\documentclass[12pt]{article}

\usepackage[letterpaper, total={7in, 9.5in}]{geometry}
\usepackage{graphicx}

\usepackage[
backend=biber,
style=ieee,
sorting=nyt,
citestyle=numeric
]{biblatex}

\usepackage[breaklinks=true,colorlinks=true,linkcolor=blue,urlcolor=blue,citecolor=blue]{hyperref}

\title{Astronomical observations: a guide for allied researchers}
\author{P. Barmby\\
Department of Physics \& Astronomy\\
University of Western Ontario\\
London, Canada N6A 3K7}

\begin{document}
\maketitle

\begin{abstract} 
Observational astrophysics uses sophisticated technology to collect and measure electromagnetic and other radiation from beyond the Earth.
Modern observatories produce large, complex datasets and 
extracting the maximum possible information from them requires the expertise of specialists in many fields beyond physics and astronomy, from civil engineers to statisticians and software engineers.
This article introduces the essentials of professional astronomical observations to colleagues in allied fields, to provide context and relevant background for both facility construction and data analysis.
It covers the path of electromagnetic radiation through telescopes, optics, detectors, and instruments, its transformation through processing into measurements and information, and the use of that information to improve our understanding of the physics of the cosmos and its history.
\end{abstract}

\section{What do astronomers do?}

Everyone knows that astronomers study the sky. 
But what sorts of measurements do they make, and how do these translate into data that can be analyzed to understand the universe?
This article introduces astronomical observations to colleagues in related fields (e.g., engineering, statistics, computer science)
who are assumed to be familiar with quantitative measurements and computing but not necessarily with astronomy itself.%
\footnote{In present-day practice there is no distinction between `astronomy and `astrophysics'; the two are used interchangeably.}
Specialized terms which may be unfamiliar to the reader are italicized on first use.
The references in this article include a mix of technical papers and less-technical descriptive works.
Shorter introductions to astronomical observations, data and statistics are given by \cite{Kremer2017,Long2017}.
Comprehensive technical introductions to astronomical observations are found in several recent textbooks \cite{chromey2016,rieke2012,Sutton2011}.

\begin{figure}
    \centering
    \includegraphics[width=\textwidth]{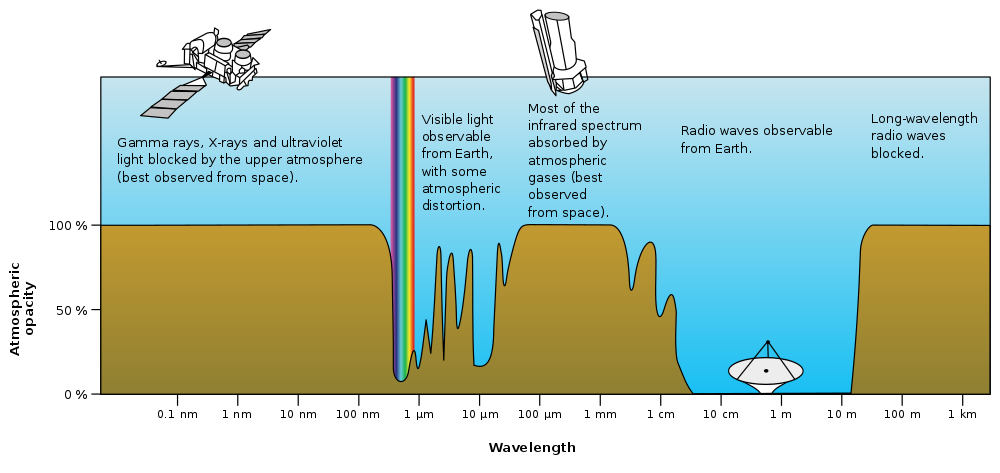}
    \caption{The electromagnetic spectrum and its transmittance through Earth's atmosphere.
    Credit: NASA, public domain 
    via \href{https://commons.wikimedia.org/wiki/File:Atmospheric_electromagnetic_opacity.svg}{Wikimedia Commons}. 
    }
    \label{fig:em_spectrum}
\end{figure}

This article focuses on astronomical observations of {\em electromagnetic radiation}. 
Electromagnetic radiation is thought of in two complementary ways: as waves characterized by {\em wavelength} $\lambda$ or frequency $\nu$, or  particles called {\em photons}, characterized by their energy (see Fig.~\ref{fig:em_spectrum}). 
Radiation with small wavelengths consists of photons with large energies, and vice versa. 
Different types of electromagnetic radiation are given different names, including ``X--ray,'' ``ultraviolet,'' ``infrared,'' and ``radio,'' but these are fundamentally the same physical phenomenon and the same theoretical understanding applies to all.

The objects that astronomers study, including stars, planets, nebulae, and galaxies,  produce radiation in different ways depending on their physical properties (e.g. composition, density, temperature) and environments.
This means that they will emit different amounts of radiation at different wavelengths.
This radiation is modified on its way to Earth by interaction with intervening gas and dust: {\em interstellar} material found
between stars but within galaxies, and {\em intergalactic} material found between galaxies.
Interaction processes include absorption, in which radiation imparts energy to matter, and scattering, in which the radiation changes direction. 
For objects well outside our own Milky Way galaxy, radiation is also {\em redshifted} --- that is, shifted to larger wavelengths or smaller frequencies --- by the expansion of the universe \cite[for an introduction see][]{Coles2001,Ferreira2007}.

Measuring the radiation from astronomical objects and interpreting those measurements is what observational astrophysicists do.
Developing physical models to predict and explain the radiation detected from astrophysical objects is the domain of theoretical astrophysicists.
For objects which radiate in all directions (most but not all astrophysical objects), the received intensity decreases with the square of the distance from the source.
Only a tiny fraction of the radiation from an astrophysical object is aimed in our direction, and that fraction is smaller for more distant objects. 
Astronomical observers are very often working in the low signal-to-noise regime, at the very edge of detectability.

Astronomical observations are nearly always {\em passive}---we have no ability to directly manipulate or experiment with the objects of interest.
In most cases we rely on radiation emitted from these objects reaching our telescopes.
This is in contrast to {\em active} remote sensing, such as sonar or radar, where radiation is transmitted to the object and scattered or reflected back for detection.
Active sensing beyond Earth is confined to radar studies of objects within the solar system: objects beyond the solar system are simply too far away for a signal to return in a detectable way or in a reasonable period of time!
Direct physical contact with the object of interest occurs in only a few situations within the solar system.
Most commonly, meteorites fall to Earth from space;
spacecraft missions have also yielded a few samples returned from the surfaces of solid bodies or particles collected from the solar wind.
Rather than ``observational astronomy,'' this kind of study would usually be called ``planetary science'' (meteorites and solid bodies) or ``space physics'' (solar wind).

Astrophysics is unique among sciences in the range of size scales involved.
It explores relationships between the largest and smallest scales, 
from the observable universe to the smallest subatomic particles.
Both the technology used to make observations, and the physics used to interpret them, span a similarly broad range.
Observational astrophysicists need to be familiar with a variety of experimental, statistical and computational techniques and technologies.

\section{Telescopes and optics}

The fundamental measurement that an astronomer makes is the amount of radiation from the sky, as a function of direction, time, wavelength or frequency, and polarization.%
\footnote{
While polarization can be quite important for understanding certain types of astrophysical objects, it's quite difficult to measure and mathematically complicated so won't be discussed further here.}
The long history of astronomical measurements began with observations made by the human eye and brain, 
later aided by architectural constructs like Stonehenge or devices like the sextant \cite{lev2013}. 
The invention and adoption of the telescope transformed astronomy.
General-level introductions to telescopes can be found in \cite{cottrell2016,graham-smith2016}. 

The most important purpose of astronomical telescopes is to act as a `light bucket."
In a rainstorm, a bucket with a larger top opening will collect more rain.
Similarly, the rate at which a telescope can collect photons from a given direction in the sky depends on the diameter of its main mirror or lens (its {\em aperture} size $D$),  and is proportional to $D^2$.
As the number of photons detected from an astronomical source increases, the uncertainty of the corresponding measurement decreases.
Because astronomical objects are far away, only a small amount of their radiation reaches us; every last photon can be important.
Using larger telescopes allows us to collect more photons, so we can detect fainter objects more quickly, or more finely subdivide (e.g. in time or wavelength) the light received from brighter objects. 

The second major purpose of an astronomical telescope is to precisely determine the location in the sky from which radiation is emanating.
Focusing the radiation can be done in one of two ways: {\em refraction} or {\em reflection}.
Refraction is familiar from other optical instruments such as eyeglasses and microscopes, and involves bending of light through a lens material (usually glass).
Very large lenses are heavy and can only be supported around their thin edges. 
Most modern large astronomical telescopes focus light using reflection by curved (usually in the shape of a conic section) mirrors that can
be supported from the non-reflecting side. 
These mirrors can be made of materials similar to familiar everyday mirrors---silver- or aluminum-coated glass---or rather different, such as the beryllium mirrors on the James Webb Space Telescope (JWST)%
\footnote{Astronomers enjoy acronyms. See \url{https://www.cfa.harvard.edu/~gpetitpas/Links/Astroacro.html}.}
or the wire surfaces of a radio telescope. 

\begin{figure}
    \includegraphics[width=5cm]{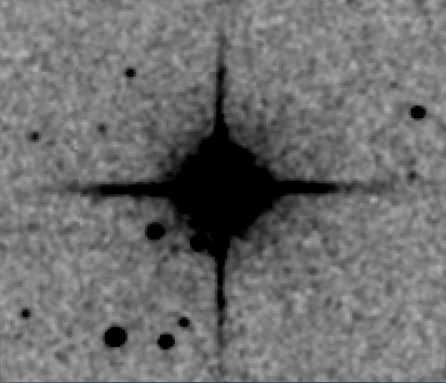}
    \includegraphics[width=12cm]{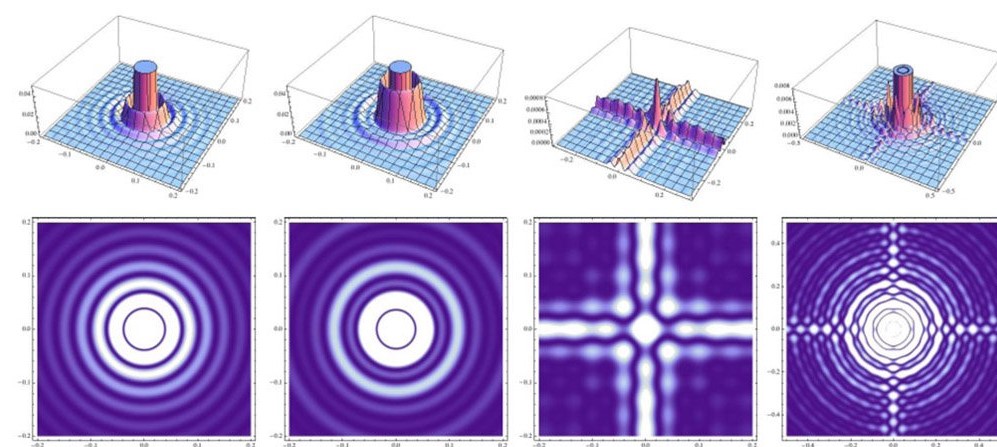}
    \caption{Point sources in astronomical images. 
    Left: Image including one bright and several fainter point sources as seen by the 2MASS telescope. 
    The images use reverse grayscale, a standard display convention in astronomy.
    All of the stars have the same shape on the image, that of the point spread function (PSF); the bright star looks larger because the faint outer parts of the PSF are visible above the background sky noise.
    The four spikes on the bright source result from diffraction of light by structures inside the telescope.    
    Right: Two-dimensional and image representations of diffraction patterns for (left to right) a finite aperture, aperture plus a central obstruction, telescope structures, and the two previous situations combined (credit: \cite{Moretto2013}, re-used with permission.) 
    }
    \label{fig:psf}
\end{figure}

A focusing telescope's ability to precisely measure the direction of radiation -- its {\em spatial resolution} -- is limited by the size of the
telescope and the diffraction of light.
Diffraction occurs when waves pass through a narrow opening or across an edge: the waves spread out and interfere with one another.
Diffraction of electromagnetic waves means that even a source of radiation of zero physical size, observed by a telescope of finite size, 
will generate an image with a finite size. 
Of course astrophysical sources have non-zero sizes, but in most cases they are far enough away to be considered points.
We call such sources {\em point sources} and the shape of their images the {\em point spread function} (PSF; Fig.~\ref{fig:psf}).
The angular size of a point source as observed by a given telescope is a way to describe  a telescope's spatial resolution, 
usually quantified by the point spread function's full-width at half-maximum (FWHM). 
In general, detecting details of an object on image scales smaller than a telescope's spatial resolution isn't possible,%
\footnote{`Super-resolution' methods used to achieve sharper images in remote sensing or microscopy are only occasionally used in astronomy \cite[e.g.][]{Li2018}: they require a detailed knowledge of the instrumentation, and most importantly, a level of signal-to-noise which is often not available in astronomical images.}
so point sources appear as pinpricks of light with no internal structure.

\begin{figure}
    \includegraphics[width=\textwidth]{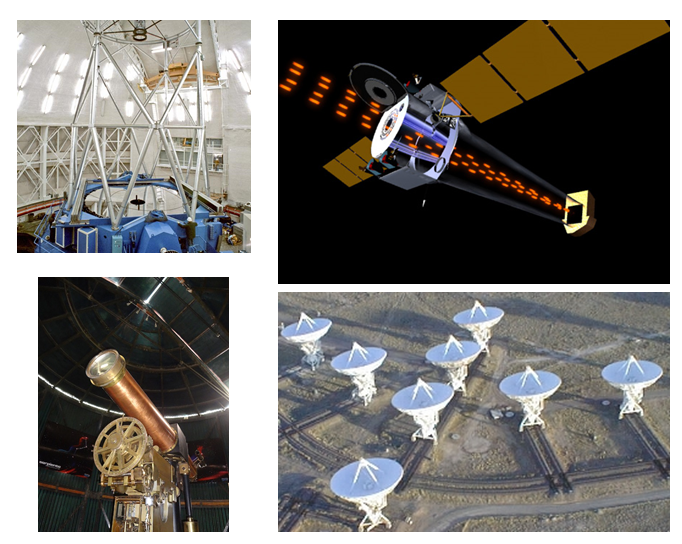}
    \caption{Astronomical telescopes. 
    Clockwise from upper left: reflecting visible-light telescope,
    reflecting X--ray telescope (note that X-rays must strike the surface at a very shallow angle of incidence to be reflected), 
    reflecting radio telescopes linked as an interferometer, 
    refracting visible-light telescope.
    Credits: Gemini Observatory/AURA,
    NASA/CXC/D. Berry,
    NRAO under license \href{https://creativecommons.org/licenses/by/3.0/}{CC BY 3.0}, 
    ``Antique Telescope at the Quito Astronomical Observatory'' - Image by D.A. Kess under License \href{https://creativecommons.org/licenses/by-sa/4.0/legalcode}{CC BY-SA 4.0}
    via 
    \href{https://commons.wikimedia.org/wiki/File:Antique_Telescope_at_the_Quito_Astronomical_Observatory_002.JPG}{Wikimedia Commons}.
    }
    \label{fig:telescopes}
\end{figure}

The spatial resolution of a telescope depends on the ratio of its aperture size to the wavelength of the radiation used ($\lambda/D$, where a smaller value is better).
At a given wavelength, a larger telescope will have better spatial resolution, but as wavelengths get larger, telescopes need to be larger to have the same spatial resolution.
For example, radio telescopes commonly observe at a wavelength of 21~cm. 
To achieve the same spatial resolution at 21~cm as a visible-light telescope working at 600~nm, a radio telescope needs to be 350,000 times larger!
Fortunately, as wavelengths get larger, telescope surfaces can be less precise.
The shape of a telescope surface has to be smooth to a fraction of the wavelength of the radiation:
visible-light telescopes must be made of very carefully polished material, but the surface of a radio telescope can be much rougher.
This makes it feasible to build large radio dishes and is why the currently-largest telescopes are those that work at radio wavelengths.

{\em Interferometers}, such as the well-known Jansky Very Large Array (Fig.~\ref{fig:telescopes}), combine the radiation received by multiple telescopes \cite{Kellerman2001}.
This achieves the spatial resolution of a telescope equivalent in size to the distance $D$ between the furthest-separated elements of the array. 
(An interferometer's collecting area is the sum of the areas of the individual telescopes, so an interferometer does not have the light collecting power of the equivalent-sized filled-aperture telescope.)
Interferometers can work at many wavelengths, but radio-wavelength interferometers are the most common.
This is both because they are more practical to construct than extremely large single radio telescopes, and because combining radiation at radio wavelengths is technically simpler.

Some specialized types of telescopes do not directly focus the incoming radiation.
The very high energies of gamma rays make it very difficult to change their direction. 
In current gamma-ray astronomy, localization of gamma ray photons is done via other techniques, including the use of coded-aperture masks, which cast a gamma-ray shadow onto a detector, or tracking the shower of visible-light Cerenkov radiation that results when gamma rays pass through the Earth's atmosphere \cite{Caroli1987,Ong1998}.
Some radio telescopes, particularly those working at low frequencies, make use of dipole antennas instead of parabolic dishes.
In these telescopes the radiation localization happens when the incoming radiation is combined electronically, trading some of the complexity of building large dishes for more complex back-end hardware and software \cite{sclocco2012}.

\begin{figure}
    \includegraphics{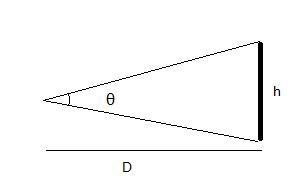}
    \includegraphics[width=8cm]{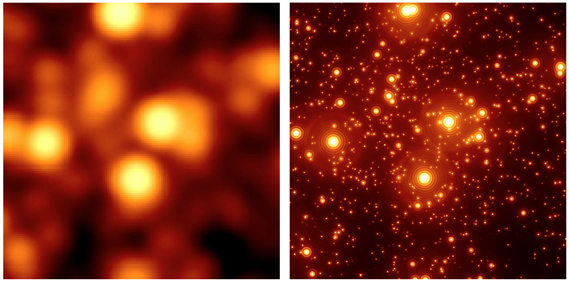}
    \caption{Angular size and angular resolution. 
    Left: in astronomy, angular size $\theta$ is easily measured while physical size $h$ and distance $D$ are usually unknown. 
    Right: the effects of improved angular resolution shown by simulated images of the same star cluster without (center) and with (far right) adaptive optics.
    Credits: N. Lantz, public domain 
    under \href{https://creativecommons.org/publicdomain/zero/1.0/deed.en}{CC0 1.0}
    via \href{https://commons.wikimedia.org/wiki/File:Angular_Size.jpg}{Wikimedia Commons},
    Giant Magellan Telescope --- GMTO Corporation.
    }
    \label{fig:ang_size_resolution}
\end{figure}

In astronomical measurements the distance to the objects being observed is generally unknown.
This means that we can only measure sizes in terms of {\em angular size} $\theta$, the fraction of a circle measured in angular units (Fig.~\ref{fig:ang_size_resolution}).
The Moon and Sun both have angular diameters of about half a degree.
The degree is rather large compared to the angular size of most astronomical objects, so smaller units are also commonly used, for example the {\em arcsecond}, which is 1/3600$^{\rm th}$ of a degree.%
\footnote{If 1/3600$^{\rm th}$ reminds you that a second is 1/3600$^{\rm th}$ of an hour, this isn't a coincidence. 
The once-a-day rotation of the Earth closely ties astronomical coordinate systems into measurements of time; for more on this concept, see Section~\ref{sec:cs_atm}.}
The angular size of the planet Saturn as seen from Earth averages about 17 arcseconds, with some variation as the planets' relative distance changes with their orbits about the Sun.
Galaxies seen halfway across the visible universe are about $10^{12}$ times larger in radius than Saturn and are further away by 
roughly the same factor; they have an angular size of a few arcseconds.
The spatial resolution of a ground-based visible-light telescope is roughly one arcsecond.
Many astronomical sources of radiation have angular sizes far smaller than this: the nearest star to the Sun, Proxima Centauri, has an angular size of about 1 milli-arcsecond.
A object whose angular size is smaller than the resolution of the telescope being used is described as being {\em unresolved}.

\section{The Earth and its atmosphere get in the way: observatories and the sky}
\label{sec:cs_atm}

Observations with telescopes located on the Earth's surface are affected by the atmosphere.
Turbulence induces refraction of radiation which slightly changes its direction.
The net effect of these slight changes is to blur images of astronomical sources such that their angular sizes are larger than they would be without the atmosphere. 
This phenomenon, called {\em seeing}, sets the spatial resolution of ground-based infrared and visible-light telescopes with sizes larger than about 10~cm.
Seeing is quantified by measuring the full-width at half-maximum of a point source
The blurring caused by seeing not only makes point sources appear larger in angular size, it also spreads out the light from objects with angular sizes larger
than the spatial resolution.
This reduces the signal-to-noise of measurements and the ability of telescopes to see detail.

Seeing is affected by weather and airflow over the surface of the Earth, so it varies with location: 
professional telescopes are built in locations where climate and topography combine to yield very good seeing, such as Maunakea on Hawai'i or the Atacama desert in Chile.
Seeing also varies with time and direction in the sky, often on a minute-by-minute basis. 
Space telescopes are outside the Earth's atmosphere and unaffected by seeing. 
This is a major reason why the Hubble Space Telescope \cite{zimmerman2010} has been such an important facility over its now more-than-20-year lifetime.
{\em Hubble\/} is not the biggest visible-light telescope, by far, nor is it closer to astronomical objects in any meaningful way, but being above the atmosphere improves its spatial resolution which is incredibly useful.

Overcoming the effects of the Earth's atmosphere can be done to some extent with {\em adaptive optics}.
This technology involves monitoring the seeing by rapidly measuring the shape and size of a reference point source, then compensating for it using a deformable mirror that changes the path of light within the telescope to compensate for what the atmosphere is doing (see Fig.~\ref{fig:ang_size_resolution}).
Bright stars can be used as references, but because seeing varies with direction, that limits the objects whose images can be corrected to those near a bright star on the sky.
It's also possible to create {\em artificial stars} by bouncing a laser beam off the Earth's upper atmosphere \cite{Wizinowich2006}; this allows adaptive optics to be used over a greater fraction of the sky.

The other major effect of the Earth's atmosphere on astronomical observations is to entirely block some wavelengths of light.
The atmosphere is transparent only to visible light, radio waves, and some wavelengths in the ultraviolet and infrared regions of the spectrum (Fig.~\ref{fig:em_spectrum}).
The transparency of the atmosphere also varies with altitude and (at some wavelengths) the amount of water vapour, meaning that high, dry sites are favorable.
Many types of radiation cannot pass through the atmosphere: detecting, for example, X--ray or ultraviolet radiation from astronomical objects requires a telescope in space.
Most space telescopes are in orbit around the Earth, but some telescopes need to be not only {\em above} the atmosphere, but {\em away} from it. 
The Earth and its atmosphere absorb radiation from the Sun and emit copious amounts of infrared radiation, making it very difficult for even space telescopes in Earth orbit to detect faint infrared radiation from distant objects.
The solution is to move such telescopes away from Earth to an orbit around the Sun, often at the L2 {\em Lagrange point} where the
Earth's and Sun's gravitational pulls are equal. 

\begin{figure}
    \centering
    \includegraphics[width=10cm]{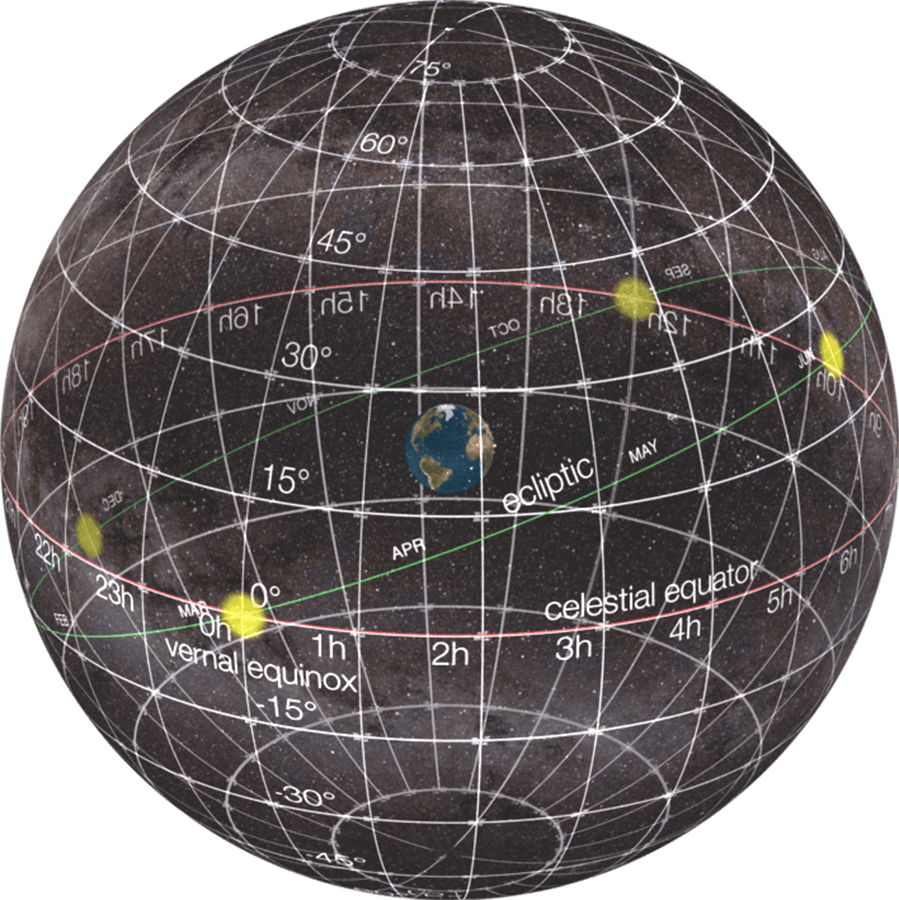}
    \caption{Astronomers visualize sky positions as locations on the ``celestial sphere'' where {\em right ascension and declination} are the equivalents of longitude and latitude on the Earth.
    Objects beyond our solar system change their positions only very slowly, while objects within the solar system appear to change positions due to their own orbits and that of the Earth (e.g. the Sun's path through the sky, known as the ``ecliptic'' and shown as a green line.) 
    Observers on Earth see different portions of the celestial sphere depending on location and time of year.
    Credit: 
    ``Celestial Sphere - Full'' - Image by C. Ready under License \href{https://creativecommons.org/licenses/by-sa/4.0/legalcode}{CC BY-SA 4.0}
    via 
    \href{https://commons.wikimedia.org/wiki/File:Celestial_Sphere_-_Full_no_figures.png}{Wikimedia Commons}.
    }
    \label{fig:cel_sphere}
\end{figure}

Infrared light from the Earth and its atmosphere is not the only kind of ``light pollution'' affecting telescopes.
From the Earth, the infrared sky is bright all the time while the visible-light sky is only bright during the day, due to sunlight scattered in the atmosphere.
A visible-light telescope on the Moon, for example, could observe with the Sun above the horizon because the Moon has no atmosphere to scatter sunlight.
However, even from the Moon or space the Sun's brightness compared to all other sources of radiation means that it's nearly impossible to observe objects which appear close to the Sun in the sky.
Objects which are located along the Sun's apparent path through the sky due to the Earth's orbit-- the {\em ecliptic}---are observable only at certain times of year (Fig.~\ref{fig:cel_sphere}).
Similarly, objects close to the Moon in the sky are also not usually observable, and some types of observations by Earth-based telescopes are also affected by scattered moonlight and can't be done when the Moon is visible.
Unlike visible-light and infrared telescopes, radio telescopes on Earth can also observe during the day and through clouds, at least at some wavelengths.

A third type of light pollution, and the usual use for the term, is the kind generated by humans.
Satellites, spacecraft and aircraft make occasional appearances in astronomical observations but are usually only a nuisance.
More significant is the effect of human-generated visible light sources on the ground \cite{kyba2018}.
These sources increase the sky background brightness and
their encroachment is one reason why professional telescopes are often built in remote locations.
Human-generated radio emission makes radio astronomy completely impossible at many wavelengths \cite{An2017}; special protected bands have been established at particularly important wavelengths (e.g., emission from atomic hydrogen in the Milky Way).
Removing the effects of radio-frequency interference (RFI) on astronomical observations is a substantial effort which is expected to become more important as new, larger telescopes come online and the use of wireless communications technologies increases.

While some telescopes can observe through the Earth's atmosphere, the types of telescopes we are discussing here can't see through the Earth, meaning that they can only observe objects which are above the horizon.
As discussed above, the location of the Earth in its orbit determines which part of the sky is in the opposite direction to the Sun and therefore visible at night. 
Location of an observatory on Earth determines which portion of the night sky is visible: for example, an observatory at the South Pole can never observe the north star because the Earth is always in the way (see~Fig.~\ref{fig:cel_sphere}). 
An observatory at the equator can see nearly the whole sky over the course of one year;
only small regions near the north and south celestial poles are never high enough to observe.
However, unfavorable weather conditions mean that few major observatories are located on the Earth's equator:
clouds and precipitation prohibit most ground-based astronomical observations!
Observatories located in space don't have to contend with rain or snow, but ``space weather,'' including magnetic activity on the Sun and micrometeroid damage from meteor streams, is occasionally an issue.

\section{Capturing the light: instruments and detectors}

Telescopes collect radiation and focus it, but this is only the first part of making an astronomical observation.
The next part is measuring that radiation: how much? what wavelength? where? when? 
Once radiation is collected by a telescope, it is directed to one or more {\em instruments} which process the incoming radiation,
for example by selecting only certain wavelengths, and direct it to a detector which records the signal.
Human eyes are not the detector of choice for professional astronomical telescopes:
electronic detectors are much more efficient and easily-calibrated collectors of radiation.
They can be made to respond to radiation at wavelengths other than the visible light we see with our eyes or the infrared radiation we feel as heat on our skin.
Many telescopes use multiple instruments, each specialized for different purposes, with new instruments often built to take advantage of advances in
detector or other technologies.

The first step in measurement is detection, that is, ensuring that a ``signal" (radiation from an astronomical object) is statistically unlikely to be a random fluctuation of noise (radiation from the telescope or instrument, a natural background such as the atmosphere or Milky Way galaxy, or human-generated interference).
Astronomical observations are very often signal-to-noise limited and carefully estimating the noise so as to evaluate the significance of a detection is a common procedure.
Astronomers often work with what statisticians call {\em censored} observations, where the precise value for a quantity cannot be measured but can be constrained to be below some quantitative limit.
Unlike in many scientific fields, such ``non-detections" are considered suitable for publication!

\begin{figure}
    \includegraphics[width=\textwidth]{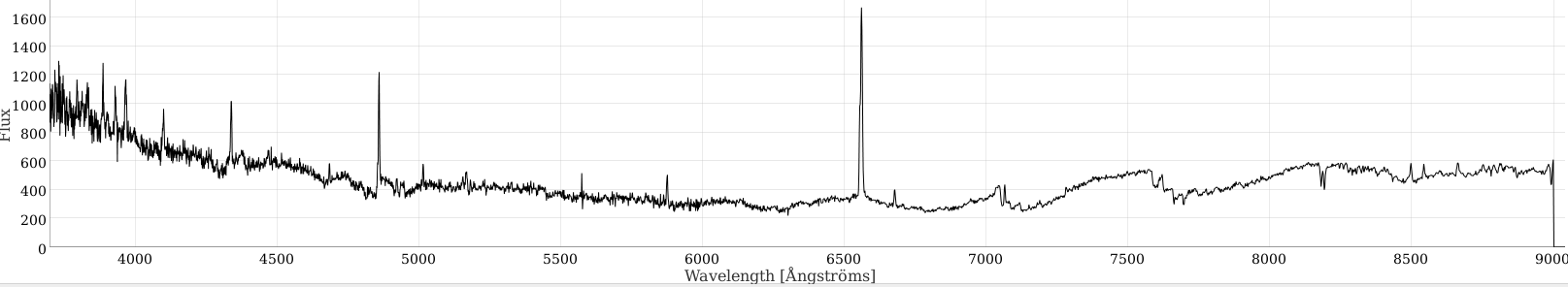}
    \includegraphics{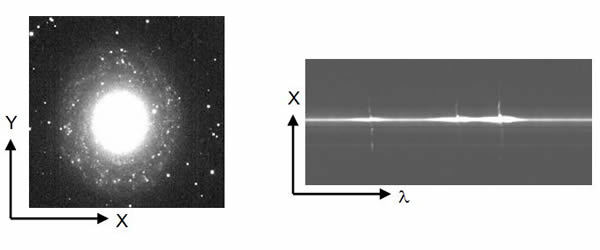}
    \includegraphics[]{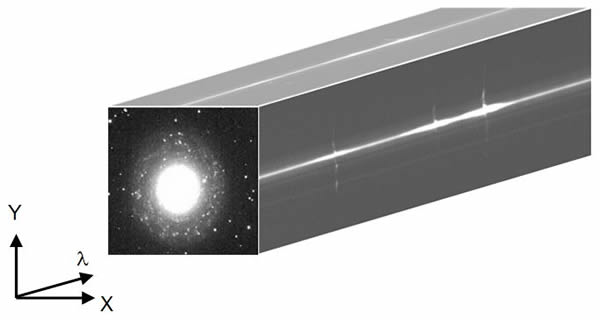}
    \caption{Astronomical measurements: 
    Top: one-dimensional spectrum of a star taken with the LAMOST telescope, from \cite{skoda2016}. 
    Center: notional two-dimensional image and spatially-resolved spectrum, 
    bottom: notional three-dimensional datacube, both from National Astronomical Observatory of Japan.}
    \label{fig:data}
\end{figure}

When faint sources of radiation are being observed, signal must be collected over a long period of time.
As with conventional photography, this period is called the {\em exposure time}, but where photographic exposure times are usually fractions of a second, astronomical exposure times can be minutes or hours.
Exposures of more than a fraction of a second require a telescope to be {\em guided} to offset the apparent motion of celestial objects due to the telescope's position on a rotating, orbiting Earth or an orbiting spacecraft.
Exposure times $t$ are chosen depending on the rate at which photons are expected to be received from the source and background ($S$ and $B$, respectively).
Usually we want to measure $S$ and estimate its uncertainty: the ratio of these two quantities is the signal-to-noise ratio (SNR) and can be shown to be 
\begin{equation}
{\rm SNR} = S/\sqrt{(S+B)/t}.
\end{equation}
A key point is that signal-to-noise increases as $\sqrt{t}$, meaning that increasing the SNR of a detection by a factor of two requires observing for four times as long. 
The equation above assumes no uncertainty in background measurement or other systematic uncertainties, which is almost never the case in practice.

The simplest type of measurement is {\em imaging}: as in everyday cameras, radiation from the telescope is focused on a {\em detector}.
In the earliest days of astronomical imaging, visible-light detectors were photographic film or plates, but now visible-light and essentially all astronomical detectors are semiconductor-based electronic devices.
The details of the detectors vary with the wavelength involved: for example, X--ray detectors can often measure the energy and arrival time of each incoming photon. 
Ultraviolet, visible-light, and infrared detectors can't do this; cameras working in these wavelengths use {\em filters} such as coloured glass to select only certain wavelengths, and a series of images to measure brightness over time. 
Detectors can have from one to hundreds of millions of picture elements, or pixels (see Fig.~\ref{fig:data}), depending on many factors including the optics of the telescope and limits of the detector technology. 
The Megacam instrument on the Canada-France-Hawaii Telescope, commissioned in 2003, has 340 megapixels and produces a typical raw data volume of $\sim 100$~GB per night \cite{Ho2018}.
The camera on the future Large Synoptic Survey Telescope (LSST) will have over 3 gigapixels and produce tens of terabytes of raw data per night \cite{Ivezic2016}.

Radio astronomy differs from visible-light and infrared astronomy in how detectors work.
At wavelengths of about 1~mm or larger, detection is {\em coherent}, meaning that the phase of the incoming radiation is measured by electronically combining the astronomical radiation with locally generated radiation of known frequency. 
(At smaller wavelengths we say that detection is {\em incoherent}.)
The fundamental reason for using coherent detection at long wavelengths is to reject thermal noise, which would otherwise dominate any astronomical signal: the thermal noise is incoherent so is not detected \cite{Wilson2013}.
Most coherent detectors have only a few pixels, so mapping an extended source requires making measurements with many pointings of the telescope, or reconstructing the sky brightness distribution with interferometry.
Technological limitations mean that long-wavelength spectroscopy (see below) is feasible only with coherent detection, and interferometry is also more straightforward with coherent detection.
Because radio interferometers must store the correlated signals from all antenna pairs, densely sampled in both time and frequency domains, they produce substantial data volumes with substantial computational challenges \cite{Offringa2016}.
Typical data rates for the Jansky Very Large Array are tens of MB per second; the future Square Kilometer Array radio telescope will produce raw data at a few TB per second, or PB per day \cite{Chrysostomou2018}.

\section{Data processing and measurements}

Recording astronomical observations with digital detectors is the first step in the measurement process.
The data stream from the instruments generally must be processed to remove instrumental signatures and calibrate the measurements into physical units. 
The procedures for this are standardized to some extent, but since astronomical instruments are typically bespoke and customized to an individual telescope, each instrument has its own idiosyncracies.
Calibration of instruments can involve a combination of laboratory testing, observations of standard sources, and cross-checks with other instruments.
Factors that need to be calibrated include {\em spatial response} (correspondence between where an object appears in an image and its true position on the sky) and {\em sensitivity} (conversion of detector counts to physical units of energy per unit time such as Watts).
Both spatial response and sensitivity can depend on characteristics of the instrument or atmosphere (e.g., temperature, humidity) as well as the detected radiation (e.g., wavelength, brightness). 

Processing (often called {\em reducing}) astronomical data requires detailed knowledge of the instrumentation, understanding of a particular observation's science goals, and scientific judgment.
Some observatories, particularly facilities with many users, provide `data pipelines' to automatically reduce data from their instruments; others expect users to process their own data.
Interferometric observations in radio astronomy produce such huge data volumes that customized hardware, as well as software, is required for data processing \cite{vdv2013}. 
For scientific analysis, astronomers prefer to use open-source software so that they know exactly what is being done to their data and are reassured that the characteristics of the physical measurement are preserved.
Commercial photo editing software is typically used in astronomy only to produce images for visualization or public relations, not for data analysis. 

The result of astronomical data reduction is a `science-ready' data product with which photometric or spectroscopic measurements can be made.
This data product may be in the form of a one dimensional spectrum, a two-dimensional image or spectrum, or a three-dimensional data cube with the three dimensions being two sky coordinates and wavelength (Fig.~\ref{fig:data}).
Astronomy uses a standard file format, FITS (Flexible Image Transport System; \cite{pence2010}) which includes both science data and metadata describing the details of the observation and subsequent processing.
Astronomical measurements are generally performed with software designed specifically to deal with astronomy's unique file formats and data conventions, although software written for other purposes is sometimes adapted (e.g., AstroImageJ \cite{collins2017}).
The Astrophysics Source Code Library \cite{Allen2017} is a repository with a comprehensive listing of astronomical software.
The AstroPy project \cite{astropy2018} is an extensive community-developed, still-evolving library of analysis code in the Python language.
As with other scientific fields, development of software in astronomy is often not recognized as ``doing science'' and its practitioners may not receive appropriate career credit \cite{muna2016}.

Measurements made on an astronomical image typically fall into two categories: {\em photometry} and {\em astrometry} \cite{Stetson2013,Vallenari2018}.
Photometry involves measuring the energy received at the telescope, either from individual objects or from extended regions within the image. 
Photometry can be either relative (to other objects in the image) or absolute (in physical units such as W~m$^{-2}$).
Absolute photometry requires calibration of the measurements, usually done by comparison to objects of known brightness.
Astrometry involves measuring the location of objects on the sky and can also be relative or absolute.
Some astronomical objects change position and/or brightness with time, for example asteroids in our solar system or pulsating stars. 
Measuring those changes requires recording a series of images at different times, with the time intervals between images matched to the expected timescales of change.
Timescales relevant to astronomical phenomena range from milliseconds to centuries, depending on the object type, and this hints at major challenges in both data acquisition and data management. 
The LSST project, one of many facing these challenges, is leveraging the experience of large particle-physics experiments to develop the necessary sophisticated computational architecture \cite{Ivezic2016}.

The contents of astronomical images are different from images found in many other scientific fields.
Some astronomical images contain a single object comprised of different regions (Fig.~\ref{fig:data}, center)
while others contain many objects, which may be resolved or unresolved (Fig.~\ref{fig:ang_size_resolution}, right).
Linear features are relatively uncommon \cite[although see][]{Bektesevic2017}, and the signal-to-noise associated with individual objects is often low.
These factors may explain why astronomers have not often used computer vision techniques developed in other fields, for example in analysis of remote sensing or medical imaging data, although some recent applications can be found  \cite[e.g.][]{Gonzalez2018,Merten2017}.

\begin{figure}
    \centering
    \includegraphics[width=4cm]{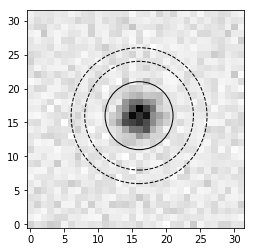}
    \includegraphics[width=7cm]{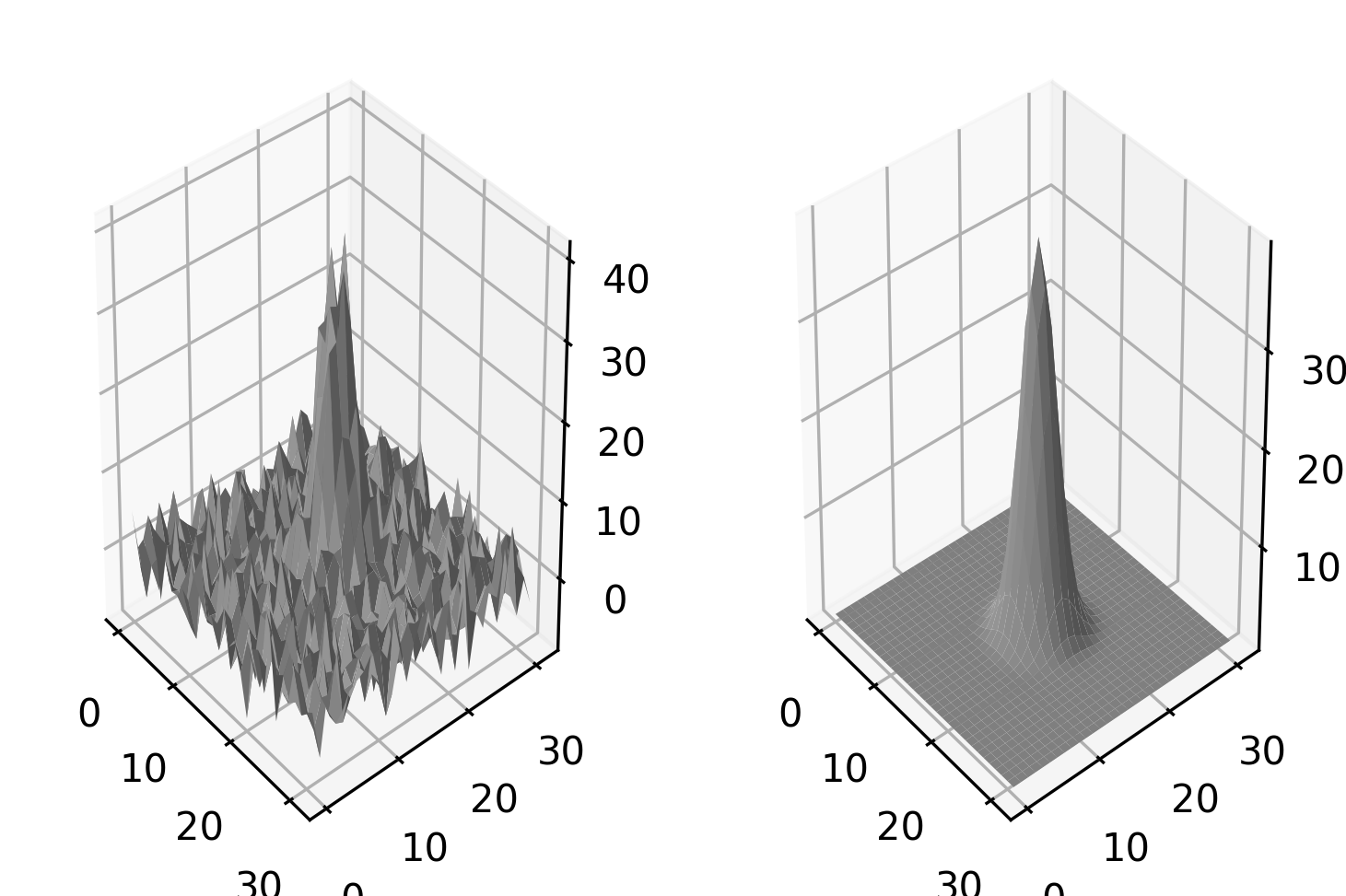}
    \includegraphics[width=4cm]{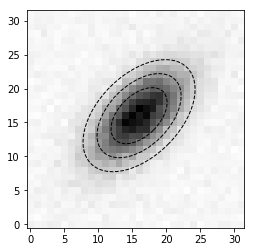}
    \caption{Astronomical photometry. 
    Left: aperture photometry measures brightness by summing the values of pixels within an aperture (solid lines) and subtracting an average background value, measured in a `background annulus' (dashed lines).
    Centre: point-spread-fitting photometry measures brightness by fitting a two-dimensional function (center right) to the pixel values (centre left). 
    Right: surface photometry measures brightness as the sum or average of pixel values within concentric circular or elliptical annuli.
    }
    \label{fig:photometry}
\end{figure}

Because modern astronomical data is, at its core, arrays of numbers, making measurements involves performing calculations on such multi-dimensional arrays.
Photometric measurements are performed on a two-dimensional image array representing brightness as a function of sky position.
Astronomical images often contain a constant or slowly-spatially-varying `background' signal, which can be due to effects within the instrumentation,
emission from the sky, or astrophysical emission, for example a star seen in projection in front of an extended gas cloud.
This background must be subtracted to correctly measure an object's brightness; depending on the type of background this
can be done on a global scale or locally near the object (Fig.~\ref{fig:photometry}).

The first step in measuring the brightness of an object is to determine its position.
This is usually done by computing the first moment (or `center of mass') of the light distribution over the relevant pixels. 
{\em Aperture photometry} involves summing the pixel values in a (usually circular) region centered on an object of interest.
Astronomical objects which are point sources should all have the same shape in the image, that of the point spread function. 
This function is often approximated as a two-dimensional Gaussian, with parameters describing the central location, width (assumed to be constant or smoothly-varying with position in an image) and height.
PSF-fitting photometry involves fitting this function over the pixels containing the object to determine the best-fit parameters;
the height of the function measures the object's brightness.
Compared to aperture photometry, PSF-fitting photometry can have improved signal-to-noise and ability to distinguish nearby objects, but it requires good knowledge of the point spread function, which is not always available.
{\em Surface photometry} applies to resolved objects such as galaxies or planets; in this case the desired measurement is the object brightness per unit area, often measured as a function of distance from some fiducial location (e.g., the equator for a planet, or the centre of a galaxy).

For small-wavelength regimes in which the energies of the incoming photons (or precise wavelengths of the incoming waves) are not directly measured by the detector, additional instrumentation is used to  disperse the radiation so that different wavelength photons arrive at different locations on a camera's detector.
This kind of instrument is called a {\em spectrograph} and the measurement of brightness as a function of wavelength is {\em spectroscopy} \cite{Appenzeller2013,Massey2013}.
Spectroscopy can be either one-dimensional, in which a single spectrum is obtained for a point-source or (part of a) resolved object, or two-dimensional, in which spectra are obtained for multiple positions within a resolved object (see Fig.~\ref{fig:data}).
As with photometry and astrometry, spectroscopy can be relative or absolute and can also be performed in a time series.
Because spectroscopy involves spreading light over more detector elements, the signal-to-noise per detector element is lower than for photometry.
This means that spectroscopy typically requires more photons to be collected, either with larger telescopes or longer exposure times or both. 
Spectroscopy is more technically demanding than photometry but the wealth of information it can provide on composition, motion, and physical conditions in astronomical objects, is incredibly valuable.

\begin{figure}
    \centering
    \includegraphics[width=10cm]{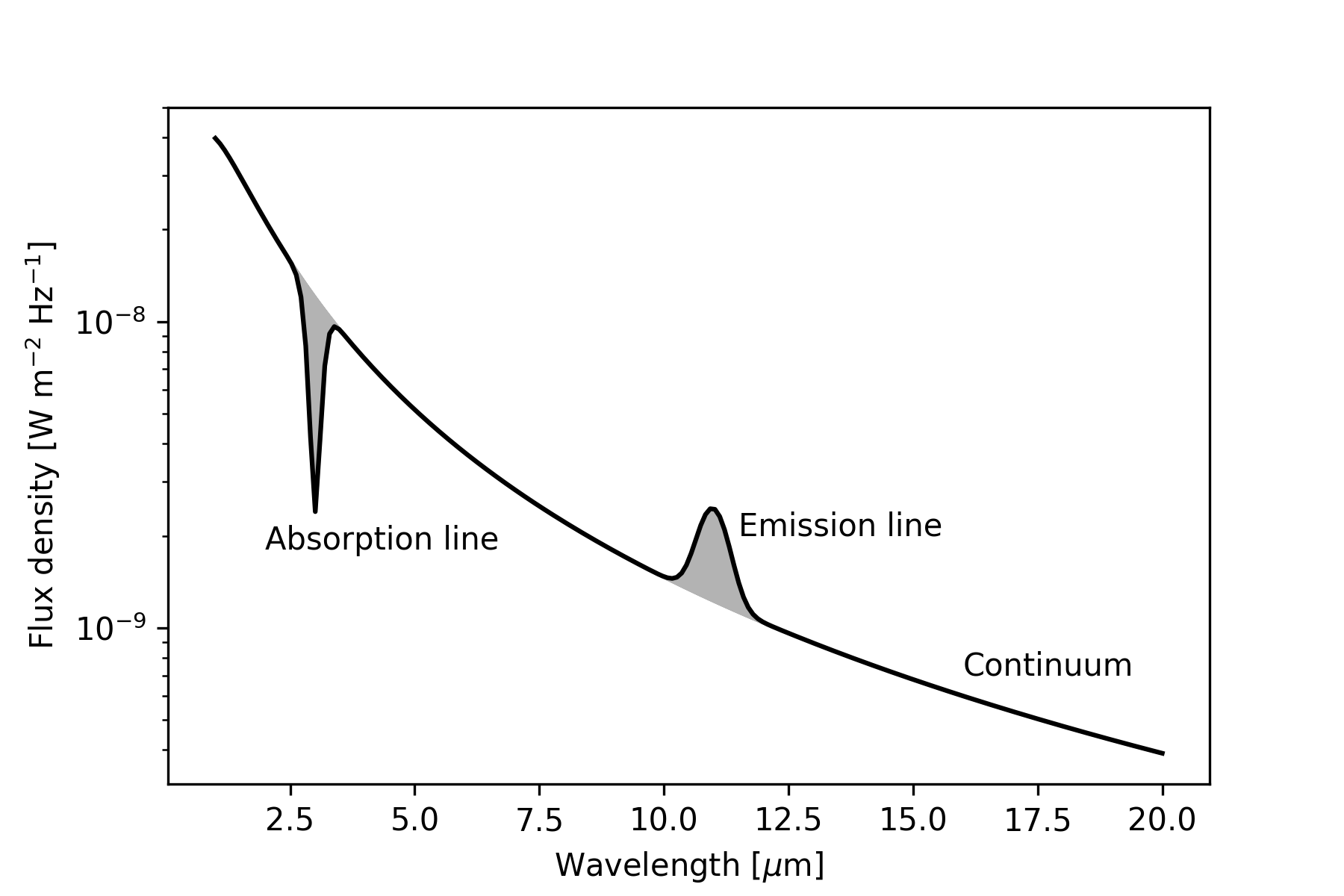}
    \caption{Notional astronomical spectrum showing absorption and emission lines and continuum. 
    Possible measurements of the lines include center, width, depth, or higher-order moments such as skew; 
    possible measurements of the continuum include its slope and/or absolute level.}
    \label{fig:spectroscopy}
\end{figure}

Spectroscopic measurements are performed on a one-dimensional array representing brightness as a function of wavelength, or on one-dimensional slices through a data cube.
Spectroscopic measurements generally fall into two categories: measurement of {\em lines} or {\em continuum} (Fig.~\ref{fig:spectroscopy}).
Spectral lines are radiation emitted or absorbed at specific wavelengths/frequencies, generated by specific atoms or molecules in the astronomical object of interest or the intervening medium. 
Line measurements can yield important physical information about conditions within an object, or its line-of-sight motion (via Doppler shift).
Identifying the source of a given spectral line is done via reference to laboratory measurements and/or theoretical calculations.
Measurement of absorption or emission lines  involves measuring their centre, width, and height either on an absolute scale or relative to the nearby continuum. 
As with PSF-fitting photometry, often the {\em line profile} is fit to a functional form that accounts for the expected response of the instrument.

Continuum radiation involves a broader range of frequencies than spectral lines and comes from {\em blackbodies} emitting thermal radiation (e.g. atoms in a stellar atmosphere) or {\em non-thermal sources} such as electrons in the magnetic field near a pulsar.
Measurement of the continuum in a spectrum involves measuring the overall shape of the spectrum over a broad range in wavelength and can be either in absolute units ({\em spectrophotometry}) or in a relative sense (e.g., measuring a {\em spectral slope}).
Spectroscopic measurements on two- or three-dimensional data extend to determining the above features as a function of spatial position.
Spectroscopic time series extend these measurements to also be a function of observation time.

How are astronomical observations used to improve our understanding of the cosmos?
While measurements may be purely descriptive, they are more likely to be used in the context of a physical model of the phenomena and processes at hand.
This may involve comparing measurements with the predictions of analytical calculations or computer simulations based on the current understanding of physics.
Many different areas of physics are relevant to astrophysics, often under conditions (e.g., temperature, density, velocity, magnetic field strength) that do not occur or cannot be replicated on Earth.
Comparison of observations with theoretical predictions often points to situations where physical understanding is inadequate, or judgment about the relevant areas of physics is incomplete.
Because astronomy is an observational science with the conditions of observation largely out of the observer's control, consideration of  selection bias and other effects in statistical inference is very important \cite{Long2017}. 

Measurements of a large number of objects within a given observation, or set of observations, are often combined into an astronomical {\em catalog}.
Catalogs have been part of astronomy for thousands of years, with some of the earliest catalogs dating from ancient China and Greece.
In their modern implementation, catalogs are published in the astronomical literature as tables within articles, if small, or made available online as databases, if large.
They can result from surveys of the entire sky, studies of a specific set of objects in a particular region, or anything in between.
Examples range from C. Messier's catalog  (1781) of just over one hundred extended celestial objects, including galaxies, nebulae and star clusters \cite{OMeara2014}, to the Two Micron All-Sky Survey catalog (2001) of over 500 million stars and galaxies detected at near-infrared wavelengths \cite{2mass}. 
While most astronomical catalogs result from deterministic data processing pipelines, new approaches to constructing catalogs based on Bayesian inference are beginning to emerge \cite{Hogg2010,Regier2018}.

Analysis of astronomical catalogs can involve characterizing the relationships between properties of the catalogued objects, measuring the joint distributions of specific properties, and searching for outliers in feature space.
To facilitate this analysis, astronomers are beginning to embrace sophisticated database methods, including the use of Structured Query Language (SQL), spurred on by the advent of the Sloan Digital Sky Survey and other large projects.
As in many other scientific fields, astronomers are rapidly expanding their use of machine learning, with topics such as classification \cite{Dick2013,Feigelson2012} and cross-identification \cite{Budavari2015} being of particular interest.
Real-time classification of time-variable events will be especially important for upcoming facilities such as LSST \cite[e.g.][]{Narayan2018}.

Allocation of telescope time for observations is performed via peer-reviewed competitive selection.
Broadly-speaking, observational programs fall into two modes: regular%
\footnote{Sometimes called PI-mode for ``Principal Investigator,'' but this is confusing terminology since surveys usually also have principal investigators.}
or survey.
In the former, the telescope is used for a specific program of investigation on one or more objects, defined and executed by a small team.
In the latter, a (typically larger) team defines a project to survey a specific area of sky or sample of objects, with the intention of making data products available for use by the team and other interested astronomers.
International collaboration is a feature of many survey programs, and the international effort to produce a sky atlas known as the {\em Carte du Ciel} \cite{Jones2000} contributed to the founding of the International Astronomical Union \cite{Trimble1997}.
Many space missions, such as WISE \cite{wise2010} or Gaia \cite{gaia2016}, were designed to perform survey programs.
The Sloan Digital Sky Survey \cite{finkbeiner2014} was the game-changing example of a fully-digital, ground-based survey.
For both types of projects, the proposing team typically retains exclusive rights to the raw observational data for a limited period of time (a few months to a few years) after which the data become publicly available through an online {\em archive}. 
Archives are incredibly valuable sources of information: the data in telescope archives can often be used for purposes other than specified in the original proposal. 
For example, more papers are now published with archival data from Hubble Space Telescope observations than with data from ``new'' proposals \cite{apai2010}.
A culture of data-sharing is well-established in astronomy but barriers remain in tapping the full potential for distribution and re-use of observations \cite{Pepe2014}.

\section{Conclusions}

Astronomical observations are a major way in which we understand the universe beyond the Earth.
The vast distances of the objects under study mean that we receive only a small amount of radiation from them, and the technical challenges involved in making measurements from this radiation and turning it into knowledge are substantial.
Maximizing the effectiveness of astronomical facilities requires state-of-the-art technology in both instrumentation and computation.

For centuries, visible light was astronomers' only source of information about the universe.
In the twentieth century, observational astronomy matured in its use of other forms of electromagnetic radiation and began to explore other messengers of the cosmos.
Particle and gravitational-wave observatories (sometimes referred to as `multi-messenger astrophysics' \cite{deangelis2018,bahcall1989,bartusiak2017}) are just beginning to yield new information inaccessible through other means.
Astronomers are fortunate to be able to expand our understanding using these methods, the next generation of electromagnetic observatories, and new approaches to understanding the information they produce.
In the past, each new technique or observational regime has yielded new discoveries, and we can hope that future new facilities will produce equally exciting results.

\section*{Acknowledgments}
Many thanks to the anonymous referee for a thorough report which helped to improve the text and presentation. 
Helpful comments from S. Gallagher, M. Gorski, A. Maxwell, S. Mittler, M. P\"ossel, and A. Rokem are also appreciated.
Any remaining errors are solely the responsibility of the author.

\printbibliography

\end{document}